\documentclass[english,twocolumn,reprint]{revtex4}
\usepackage[T1]{fontenc}
\usepackage[latin9]{inputenc}
\usepackage{graphicx}

\makeatletter
\@ifundefined{textcolor}{}
{%
 \definecolor{BLACK}{gray}{0}
 \definecolor{WHITE}{gray}{1}
 \definecolor{RED}{rgb}{1,0,0}
 \definecolor{GREEN}{rgb}{0,1,0}
 \definecolor{BLUE}{rgb}{0,0,1}
 \definecolor{CYAN}{cmyk}{1,0,0,0}
 \definecolor{MAGENTA}{cmyk}{0,1,0,0}
 \definecolor{YELLOW}{cmyk}{0,0,1,0}
}

\makeatother

\makeatother

\usepackage{babel}

\makeatother

\usepackage{babel}

\makeatother

\usepackage{babel}

\usepackage{babel}

\makeatother

\usepackage{babel}
\begin{document}

\title{Magnetic force microscopy investigation of arrays of nickel nanowires
and nanotubes}

\author{M. R. Tabasum$^{1}$, F. Zighem$^{2}$, J. De La Torre Medina$^{3}$,
A. Encinas$^{4}$, L. Piraux$^{1}$ and B. Nysten$^{1}$}

\affiliation{$^{1}$ Institute of Condensed Matter and Nanosciences \textendash{}
Bio and Soft Matter, Université catholique de Louvain, Belgium}

\affiliation{$^{2}$ Laboratoire des Sciences des Procédés et des Matériaux, CNRS-Université
Paris 13, 93430 Villetaneuse, France}

\affiliation{$^{3}$ Facultad de Ciencias Físico-Matemáticas, Universidad Michoacana
de San Nicolás de Hidalgo, Morelia, México}

\affiliation{$^{4}$ Instituto de Física, Universidad Autónoma de San Luis Potosí,
Mexico}
\begin{abstract}
The magnetic properties of arrays of nanowires (NWs) and nanotubes
(NTs), 150 nm in diameter, electrodeposited inside nanoporous polycarbonate
membranes are investigated. The comparison of the nanoscopic magnetic
force microscopy (MFM) imaging and the macroscopic behavior as measured
by alternating gradient force magnetometry (AGFM) is made. It is shown
that MFM is a complementary technique that provides an understanding
of the magnetization reversal characteristics at the microscopic scale
of individual nanostructures. The local hysteresis loops have been
extracted by MFM measurements. The influence of the shape of such
elongated nanostructures on the dipolar coupling and consequently
on the squareness of the hysteresis curves is demonstrated. It is
shown that the nanowires exhibit stronger magnetic interactions than
nanotubes. The non-uniformity of the magnetization states is also
revealed by combining the MFM and AGFM measurements.
\end{abstract}


\maketitle

\section{Introduction}

The manufacturing of nanostructures such as nanowires (NWs) and nanotubes
(NTs) is a quite rousing area in the domain of materials engineering
because of their possible applications in diverse fields such as magnetism,
catalysis, nanomedicine, information processing and magnetic recording
{[}1,2{]}. The development of ultra-high density recording medium
requires nanostructures with magnetically isolated grains. The studies
have been done to exploit the magnetic anisotropy of super paramagnetic
nanoparticles, using MFM, by aligning their magnetic moment with an
external applied field {[}3{]}. Another approach to overcome the super-paramagnetic
limit consists in using magnetic nanostructures with enhanced anisotropy.
This can be achieved by using nanocylinders instead of nanodots {[}4{]}.
Out of the numerous techniques available for manufacturing ferromagnetic
NWs and NTs, the template based growth has triggered a lot of attention
given that it is inexpensive and quite versatile {[}5,6{]}. The magnetic
characteristics of NWs have been studied from various viewpoints such
as magnetization reversal, magnetostatic interactions, microwave properties
and calculations of their intrinsic switching field distributions
(SFD) {[}7,8,9,10,11{]}.

On the other hand manufacturing magnetic NTs is more difficult compared
with NWs. It is the reason why their magnetic properties, magnetization
reversal for instance, have not been so extensively explored despite
their potential advantages over NWs, such as tunable geometry and
reduced magnetic material volume. In particular, interwire interactions
have proven to affect the magnetic properties of arrays of NWs, specifically
their magnetization reversal process and switching field distribution
(SFD) {[}11,12{]}. It has been shown theoretically that NTs exhibit
core-free magnetic configuration resulting in uniform switching fields
leading to controllable magnetization reversal process {[}13{]}. Recently,
experimental researches on magnetic NTs have become an attractive
field to be investigated {[}15,16,17,18,31{]}. The knack to tune NWs/NTs
geometries and interwire distance permit to control the magnetostatic
energies in order to get the desired magnetic properties. However,
their integration into novel devices necessitates to fully understand
their properties, in particular magnetostatic interactions. Major
hysteresis loops of M(H) curves provide basic understanding of the
magnetic properties. However, this technique alone is not sufficient
for in depth quantitative determination of the magnetic interactions
of the nanoscopic materials entities and needs complementary measurements
to assess and to gain more understanding of the magnetic properties
{[}19{]}. Magnetic force microscopy (MFM) has proven to be suitable
for the determination of the magnetization hysteresis curves at a
local scale and to gain insight into the interwire dipolar interactions
{[}10,11,20{]}.

MFM is a complementary technique to perform magnetic characterization
at a local scale and has been widely employed to study densely packed
nanoparticle assemblies and patterned media, which are of considerable
interest for applications in magnetic recording {[}11,12{]}. Yet a
more basic understanding of magnetic processes in such assemblies
requires a more clear and precise comprehension of intrinsic particle
properties and how they differ from the collective ones of densely
packed ensembles. Moreover, in very dense magnetic assemblies the
dipolar interaction is strong and local (or intrinsic) properties
of the particles are sensibly modified by this field. Therefore, low
and medium density assemblies are also very interesting from a fundamental
point of view in order to gain a better understanding of the information
that can be obtained by MFM regarding local versus collective properties,
as done for example in artificial spin ice nanomagnet assemblies{[}14{]}.

In this study, a comparative MFM study of the magnetization reversal
process of arrays of Ni NWs and NTs fabricated by electrodeposition
in nanoporous polycarbonate (PC) membranes is presented. This study
aims at presenting the MFM as a complementary technique to the bulk
magnetometry technique AGFM, for the in-depth investigation of elongated
magnetic nanostructures embedded into dielectric matrices. Considering
the same nanoporous template, the dipolar interactions is found to
be stronger in NWs arrays than in NTs arrays and hence the SFD of
the NWs is broader than for NTs.

\section{Materials and methods}

\subsection{Manufacturing of magnetic nanotubes and nanowires}

Arrays of Ni NWs and NTs have been fabricated by electrodeposition
in the same track-etched 21 $\mu$m thick PC membranes with pore diameter
($D$) of $150\pm5$ nm and packing density ($P$) of 6$\%$. Before
the electrodeposition, one side of the membranes was covered with
a metallic layer (Au), by e-beam evaporation that acts as a cathode.
A Cr layer 10 nm thick was fi{}rst evaporated to serve as adherent
layer between the template and the Au layer. Ni NWs were grown at
a constant potential of $-1.1$ V from an electrolyte containing 1
M NiSO$_{4}$$\cdot$6H$_{2}$O and 0.5 M H$_{3}$BO$_{3}$. Ni NTs
were fabricated using a two-step procedure as depicted in Figure 1a.
The process started by growing Ni/Cu core/shell NWs at a constant
potential of $-1.0$ V using a 0.4 M Ni(H$_{2}$NSO$_{3}$)$\cdot$4H$_{2}$O,
0.05 M CuSO$_{4}$$\cdot$5H$_{2}$O and 0.1 M H3BO3 electrolyte.
The Cu core was later etched by the electrochemical etching at a potential
of $+0.2$ V {[}21{]}.

\subsection{SEM Imaging}

\begin{figure}
\begin{centering}
\includegraphics[bb=20bp 140bp 750bp 580bp,clip,width=8.5cm]{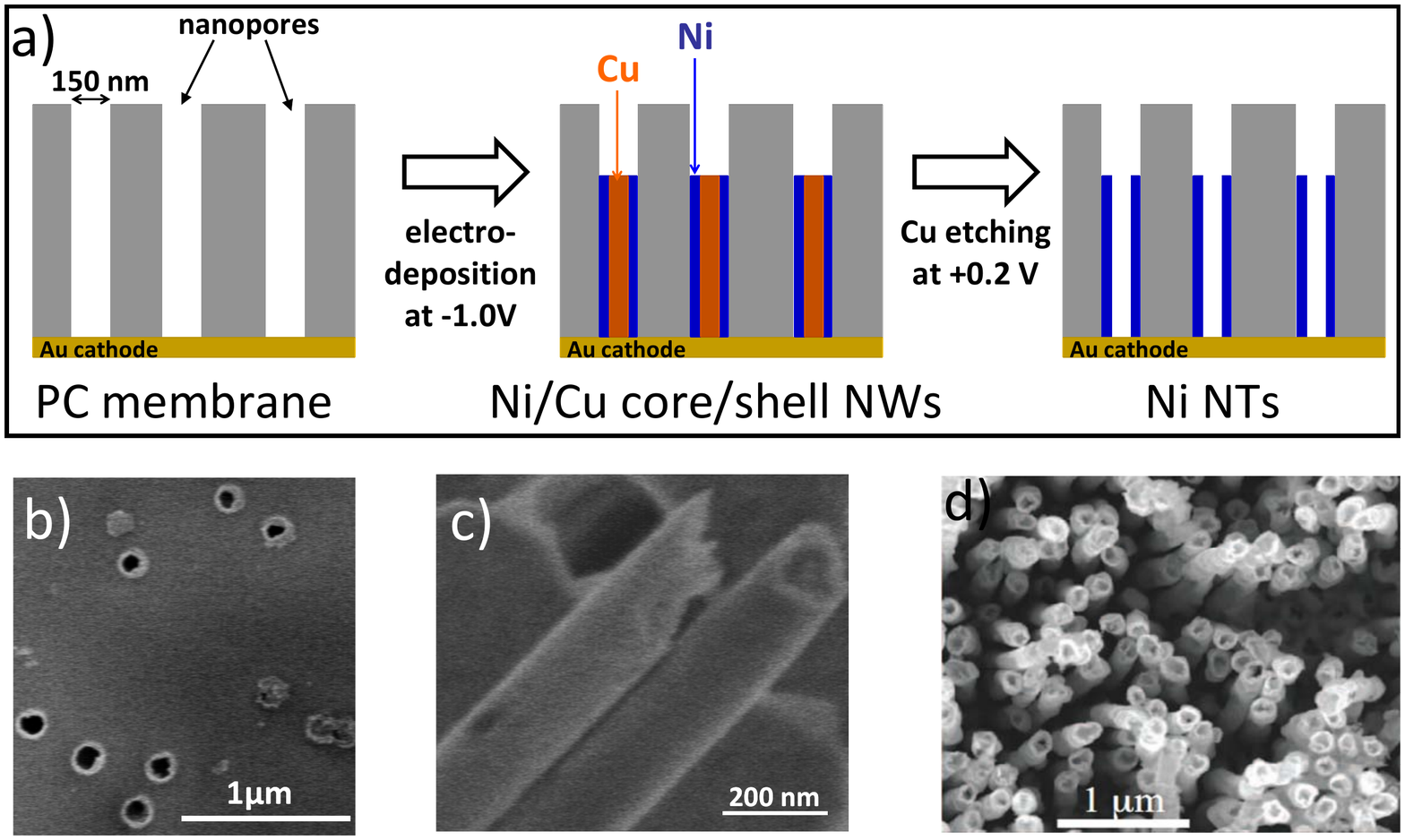} 
\par\end{centering}

\caption{a) Sketch of the NTs fabrication process presenting the two different
steps. b): SEM images of Ni NTs in the PC membrane after etching the
metal cathode (c); of two Ni NTs after the dissolution of the PC membrane
and (d) of a large number of NTs obtained with a higher packing density
membrane ($P=10\%$.).}
\end{figure}

Scanning electron microscopy (SEM) images of the NTs in the PC template
as well as of free NTs were performed using a field emission scanning
electron microscope (Leo-982 Gemini). A smooth surface where all the
nanowire tips are close to the surface on one side, allowing the microscopy
analyzes, has been obtained by removing the Au and Cr layers using
a chemical etching procedure (Fig. 1b). The Au layer was first etched
using a KI (100 g/l) + I$_{2}$ (25 g/l) solution and then the Cr
layer was removed using a Cr etching solution containing KMnO$_{4}$
(52 g/l) + 5 M NaOH (600 ml/l). Moreover, in order to show the entire
shape of the NTs by SEM, pieces of the PC membrane has been dissolved
(Fig. 1c). In the first step, about 20 to 25 dichloromethane droplets
were spread on the substrate to well dissolve the PC membrane, leaving
only the exposed NTs.

\subsection{In-field MFM experiments}

In-field MFM experiments have been performed under ambient conditions
using an Agilent 5500 microscope (Agilent Technologies) equipped with
a 100 $\mu$m closed-loop scanner, for high-precision control of the
scanner position in $x$ and $y$ by attenuating the non-linear and
hysteretic behavior of the piezo-positioners. So, in practice, we
could scan the same area during all the measurements. The MFM probes,
Asylum ASYMFMHC high coercivity ($H_{C}=5000$ Oe) (Asylum) with a
force constant around 2 N.m$^{-1}$ and a resonance frequency of about
70 kHz were used for this study. The analyzes were realized in amplitude-modulation
(AM-AFM) using a double pass procedure. First, the topography of one
line was recorded in standard intermittent-contact mode. Then, the
probe was lifted up a few tens of nanometers (typically 60 nm) and
the same line was scanned at constant probe-surface distance; the
phase signal proportional to the magnetic interaction gradient was
simultaneously recorded.

A custom-built electromagnet has been used to modify the setup of
the instrument for performing in-field MFM experiments. Before launching
the experiments, the NWs/NTs arrays were magnetically saturated along
their axis ($+Oz$) under a magnetic field of $H=+3$ kOe while the
MFM probe tip was saturated in the opposite direction ($-Oz$). This
resulted in attractive bright contrast for the NW (or NT) on the MFM
phase images. Then, for the in-field measurements, the magnetic field
was applied in the direction of the magnetization of the probes ($-Oz$)
to switch the magnetization of the NW/NT leading to the progressive
observation of dark spots. It is worth noting that, contrary to the
in-situ measurements {[}7{]}, during these in-field measurements,
the applied magnetic field was not switched off during the measurements.
This procedure was continued with incremental magnetic field until
the field was sufficient to saturate all the NWs/NTs. Moreover, the
use of high-coercivity probes allowed to avoid the magnetization reversal
of the tip in the applied field range {[}-500 Oe; +500 Oe{]}. The
MFM-based magnetic hysteresis curves were obtained by counting at
each increment of field the number of switched and unswitched NWs/NTs.

\subsection{AGFM measurements}

Finally, bulk magnetization curves were also obtained, using an alternating
gradient field magnetometer (AGFM, Lakeshore) and were compared to
the MFM-magnetization curves. The basic difference between these two
measuring techniques, i.e AGFM and MFM, is that AGFM senses the magnetization
of the whole sample which can include magnetization inhomogeneities,
partially switched magnetizations, domains, etc. while MFM gathers
local information by counting the occurrence of switching events as
a function of the applied field. The sample size was $2\times2$ mm$^{2}$
which contains a larger number of NTs and NWs. The AGFM measurements
were performed on the same samples as for MFM characterization, after
the chemical etching of the Cr and Au layers which guarantees the
same conditions for the arrays of NTs and NWs.

\section{Results and discussion}

\begin{figure}
\includegraphics[bb=30bp 165bp 330bp 800bp,clip,width=8.5cm]{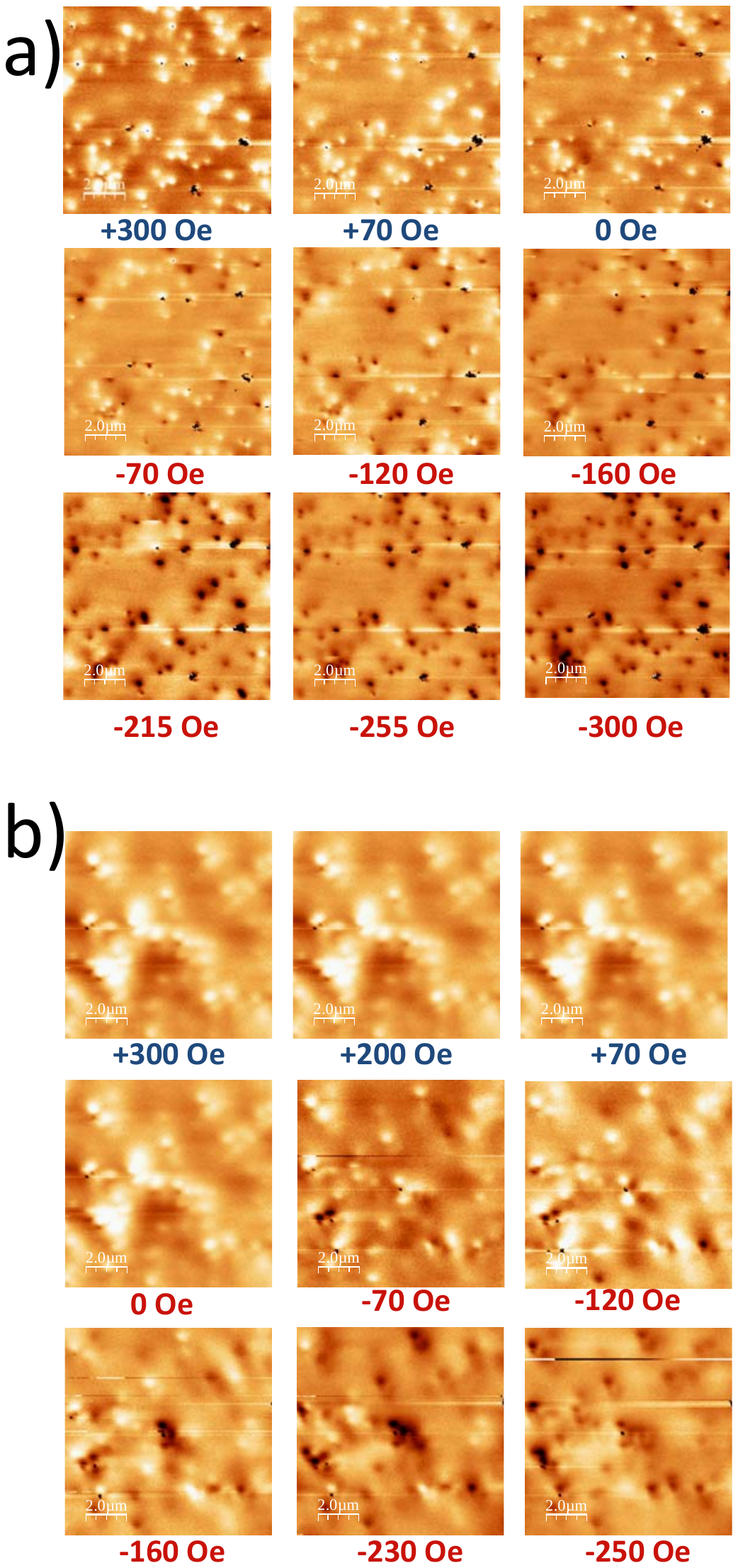} 

\caption{In-field MFM images of Ni NWs (a) and NTs (b) starting at saturation
in a positive 300 Oe field and at various negative magnetic fields
displaying the magnetization reversal progress. The images are scanned
at a fix area ($10\times10$ $\mu$m$^{2}$). The first image at top
left of each series is the topographic image of the studied area.
For the MFM images of the NW arrays at positive fields, white arrows
point spots due to artifacts (VdW interactions) and black ones point
to spots corresponding to switched NWs.}
\end{figure}

First, it is important to note that due to their different surface
area, the effective packing density ($P$) for the NTs array is less
than for the NWs array considering the same template. The effective
packing density $P$ is defined as the total area of the top end of
the NW ($S_{NW}$) (or of the NT ($S_{NT}$)) divided by the total
surface area of membrane under consideration. By combining MFM and
SEM images, the effective packing density in the NWs array is found
to be $P_{NW}\sim6\%$ Whereas, the effective packing density in NTs
array was $P_{NT}\sim5\%$ calculated using following equation.
\begin{equation}
P_{NT}=P_{NW}\left(1-\beta^{2}\right)
\end{equation}

where $\beta=\frac{r_{1}}{r_{2}}$ is the ratio of internal and external
radii. 

After etching the cathode metals (Au and Cr), the NTs embedded in
the PC membrane have been revealed as shown in Figure 1b. This surface
corresponds to the one probed by MFM. Free NTs were also analyzed
by SEM (see Figure 1c and 1d). Figure 1d has been obtained from NTs
electrodeposited in a PC membrane with a larger pore density ($P=10\%$).
From these images, the lengths of the studied NTs and NWs were found
to be around 5 $\mu$m.

Figure 2a present typical $10\times10$ $\mu$m$^{2}$ MFM images
obtained at the top surface of the NWs array while in Figures 2b,
images of the NTs array are presented. The corresponding topography
images NWs and NTs samples are presented on top left of the Figures
2a and 2b. The spot sizes (black or white) are larger than the nominal
diameters (150 nm) of the NWs/NTs because the MFM tip measures their
dispersive stray fields from the top and not their exact dimensions.
One would expect NTs to have a non-uniform MFM signal because of its
hollow inner core. Nevertheless, from all the images, only \textquotedblleft{}uniform\textquotedblright{}
white or black spots are distinguishable as a function of the applied
magnetic field. It has been demonstrated experimentally that the Ni
NWs of 250 nm diameter show single domain structure whereas of 2 $\mu$m
diameter exhibit core-shell cylindrical domains {[}34{]}. Subsequently,
the arrays of NTs and NWs have been treated as apparent bistable systems
(uniform magnetization either parallel or antiparallel to the probe
magnetization) to analyze the MFM images even if the magnetic moments
distribution could be more complex. This behavior is discussed at
the end of this section.

The second image (first rows) in Fig. 2a and Fig. 2b correspond to
a state where all the NW/NT appear uniformly magnetized in a field
$H_{0}=+300$ Oe along $+Oz$ while the tip is magnetized along $-Oz$.
Then, a series of magnetic fields along $-Oz$ were applied. The successive
switching of the NWs and NTs started around +200 Oe and 70 Oe until
the application of around 300 Oe and 250 Oe respectively. In contrast
to the NTs, the switching of the NWs started even before reaching
remanent state which results in some reversed NWs (dark spots) at
zero applied field (pointed with black arrows). In Fig. 2b the black
spots with horizontal straps from +300 Oe till 0 Oe (pointed with
white arrows) are not due to the magnetic signals but to artifacts
due to Van der Waals interactions acting on the tip that briefly touches
the sample surface. These spots do not persist once the applied field
and the tip stray field were in the same direction, Fig 2b from -70
Oe till -250 Oe.

Because only two contrasts (black or white spots) were identifiable
from the MFM images, we made the assumptions that only two states
are possible: \textquotedblleft{}up\textquotedblright{} or \textquotedblleft{}down\textquotedblright{}.
Thus, the local- or MFM-hysteresis cycle has been calculated by counting
the number of NWs (or NTs) with \textquotedblleft{}up\textquotedblright{}
states and \textquotedblleft{}down\textquotedblright{} states. The
normalized magnetization could be written as:
\begin{equation}
\frac{M^{MFM}(H)}{M_{S}^{MFM}}=\frac{n_{up}-n_{down}}{n_{up}+n_{down}}
\end{equation}

Where $n_{up}$ and $n_{down}$ are the number of NTs (NWs) with up
and down states, respectively.  $M_{S}^{MFM}$ is the effective saturation
magnetization (total number of NTs (NWs) in the image while $M^{MFM}(H)$
is the effective magnetization at an applied magnetic field $H$.
The remanence was extracted from the image recorded at zero applied
field. It is thus the following quantity:
\begin{equation}
\frac{M^{MFM}(0)}{M_{S}^{MFM}}
\end{equation}

The (MFM)-coercive field was extracted from the image where the number
of NWs/NTs with \textquotedblleft{}up\textquotedblright{} and \textquotedblleft{}down\textquotedblright{}
states is same ($n_{up}=n_{down}$). 

Bulk magnetization curves with longitudinal applied magnetic field
were also obtained by AGFM on a piece of the same sample used for
MFM measurements. Figure 3 presents a comparison between the bulk
magnetization curves for the NTs array (red continuous line) and for
the NWs array (blue dashed line). These magnetization curves clearly
reveal that the NTs curve is less tilted (more square) compared with
the NWs curve. The remanence of the NTs array is about 0.8 while it
is around 0.5 for the NWs array. Moreover, the saturation field for
the NWs is about two times higher (\textasciitilde{} 2000 Oe) than
for NTs (\textasciitilde{} 1000 Oe). These observations are, in first
approximation, coherent with the observations made by MFM where the
NWs reversal magnetization begins before and ends after the reversal
of the NTs.

\begin{figure}
\includegraphics[bb=25bp 330bp 350bp 570bp,clip,width=8.5cm]{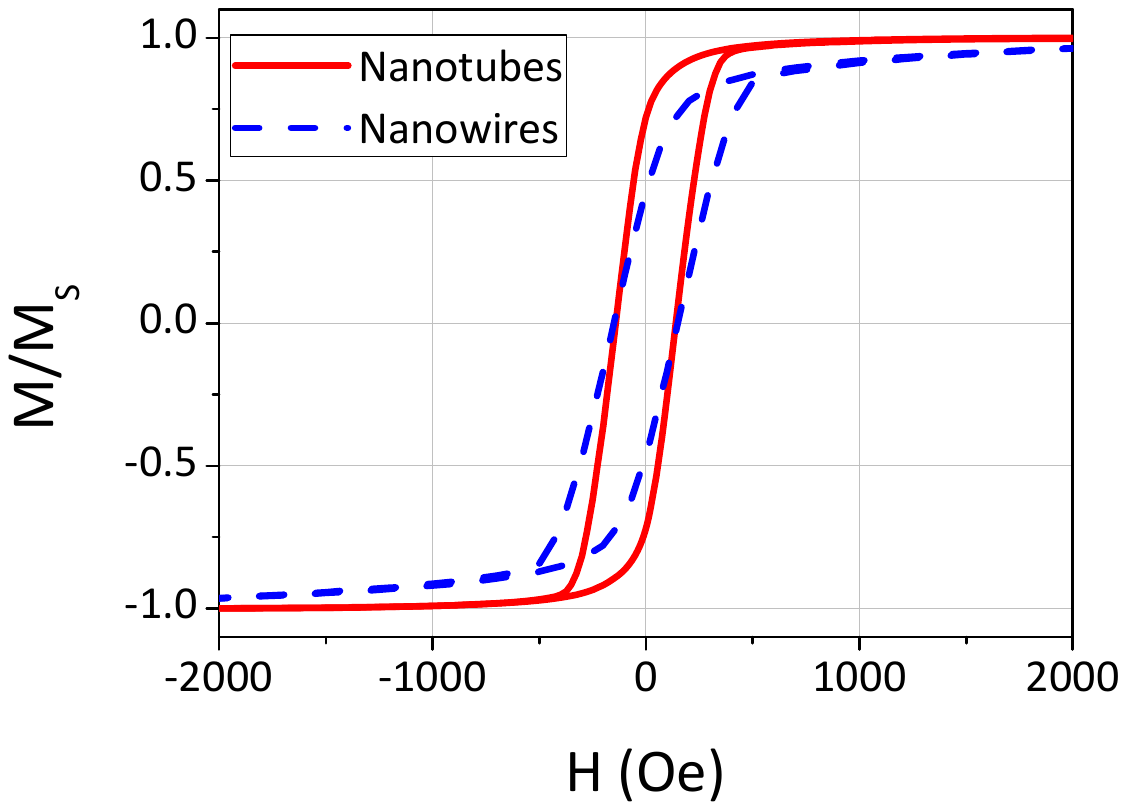} 

\caption{Normalized AGFM magnetization curves obtained with a magnetic field
aligned along the revolution axis from the arrays of NTs (red solid
line) and NWs (blue dashed line).}
\end{figure}

The difference of the two bulk magnetization curves could be understood
solely in terms of the dipolar interactions and shape anisotropy contributions,
since the magnetocrystalline anisotropy contribution, in NWs and NTs,
may be neglected {[}21,22,23{]}. Considering that the magnetization
inside each NW and NT is uniform and because both NW and NT have the
same symmetry with a large aspect ratio, the same mean fi{}eld model
used for NWs can be used for NTs, where the eff{}ective fi{}eld $H_{eff}$
in the saturated state is {[}31{]}: 
\begin{equation}
H_{eff}=2\pi M_{S}-6\pi M_{S}P_{NW}\left(1-\beta^{2}\right)
\end{equation}

where $M_{S}$ is the saturation magnetization for Ni and $P_{NW}$
is the effective packing density of the considered array. The first
term of this equation corresponds to the shape anisotropy contribution
which is same for both high aspect ratio NWs and NTs and the second
term is the dipolar interactions contribution characterized by the
effective packing density $P_{NW}$ {[}32,33{]}. It has already been
discussed that due to their different volume, the packing fractions
for NT ($P_{NT}$) is less than the NW ($P_{NW}$) which results in
a weaker interaction field and thus a higher effective field and a
strong uniaxial anisotropy for NTs. So, the key differences observed
in their hysteresis loops arise from their respective effective fi{}elds
and particularly from the dipolar interaction fi{}eld. Moreover, the
high packing value of the template results in strong dipolar interaction
and broader switching field distribution (SFD) {[}24,25{]}. The shearing
of hysteresis loops depends on the value of the dipolar interaction
field so the hysteresis loops of NWs are more sheared than NTs. 

\begin{figure}
\includegraphics[bb=25bp 100bp 350bp 570bp,clip,width=8.5cm]{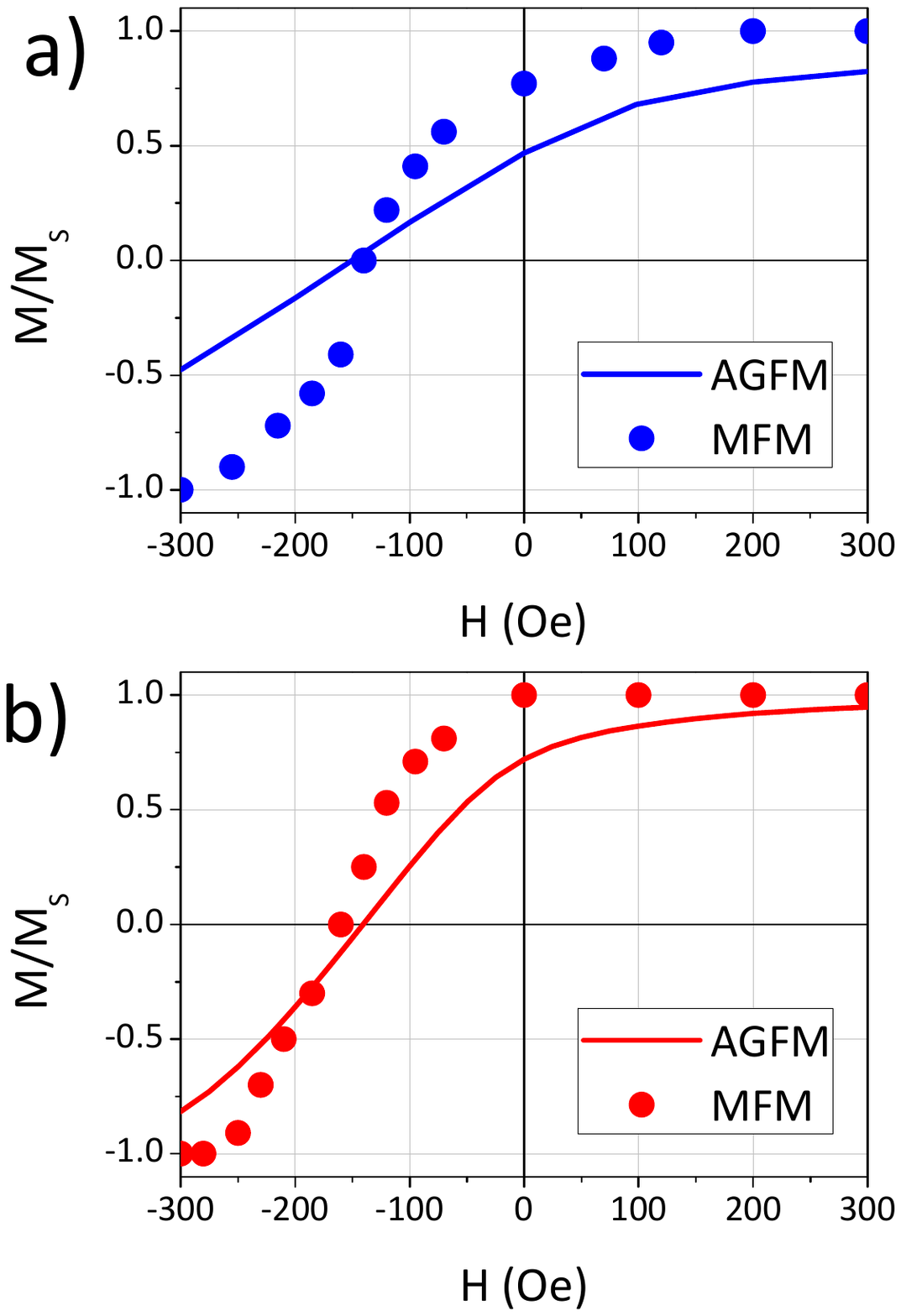} 

\caption{Comparison of the in-field MFM magnetization curves with the ones
obtained by AGFM measurements from the array of NWs (a) and NTs (b),
respectively.}
\end{figure}

Magnetometry and MFM hysteresis curves differ significantly, as presented
in Figure 4 for the array of NWs (a) and for the array of NTs (b).
The MFM hysteresis curves were obtained by analyzing the MFM images
presented in Fig. 2. From the MFM images, a remanence of $100\%$
was found in the case of NTs and around $80\%$ for the NWs while
it reaches $80\%$ and $50\%$ from the AGFM curves, respectively.
Moreover, a saturation field of 300 Oe is deduced in both investigations
for NTs while it was found to be around 1000 Oe and 2000 Oe for NWs
arrays. In contrast, the coercive field values obtained from MFM and
AGFM curves are the same.

These comparisons suggest that the NWs and NTs have non uniform magnetization
states. It has recently been shown {[}10{]} that a very good agreement
between MFM and AGFM curves is found for arrays of ferromagnetic NWs
in a state of uniform magnetization, i.e. single domain regime. It
is worth mentioning that different series of MFM images were taken
at different areas of the samples and no significant changes were
noticed on the resulting hysteresis loops. Hence, the antagonism between
the local probing technique of the MFM approach compared with the
bulk probing technique of AGFM can be excluded. In addition to being
a local technique, MFM is a surface technique where the in-depth magnetic
configuration cannot be probed whereas AGFM probe the whole volume
of the NTs and NWs arrays. 

Indeed, the diameter of the studied NWs is around 150 nm which is
far from the critical diameter below which a coherent rotation of
the magnetization is expected for infinite Ni NW ( $D_{coh}=7.3\ell_{ex}\sim54$
nm {[}22{]}) and non-uniform micromagnetic configurations could be
present. Interestingly, from Figure 4, it can be seen that the mismatch
between bulk and MFM measurements is less pronounced in the curves
for the NTs array. This finding could be explained by the specific
shape of the NTs compared to the NWs where the shell width is about
20 nm which prevents the presence of radial domains. Therefore, from
these comparisons, one can notice that in both cases, the micromagnetic
configuration is certainly non uniform. To reinforce this argument,
micromagnetic simulations on an isolated NT and an isolated NW have
been performed. The calculations have been performed by using the
NMAG package {[}33{]} and the results are presented in Figure 5.

The dimensions used during simulations for the NW are: length $L=5$
$\mu$m and diameter $D=150$ nm while for the NT $L=5$ $\mu$m,
outside diameter $D_{out}=150$ nm and inside diameter $D_{in}=130$
nm. Note that these geometrical dimensions are close to the experimental
ones. The magnetic parameters: magnetization saturation $M_{S}=0.48\times10^{3}$emu.cm$^{-3}$
and exchange stiffness $A=1\times10^{-6}$erg.cm$^{-1}$ correspond
to bulk values for Ni material. Indeed, it has been shown that the
electrodeposition method seems not to modify the intrinsic magnetic
parameters {[}35,36{]}. The magnetic moments distributions at remanence
(after saturating the NW and the NT along the revolution axis) are
presented in Figure 5. It clearly depicts that the distribution at
zero applied field is non uniform in both cases. 

It is worth noting that contrary to the experimental hysteresis cycles
(Figure 3), the calculated hysteresis cycles (Fig. 5) are close one
to each other (the coercive field of the NT is slightly weaker than
the NW one) which strongly endorses that the main difference appearing
between the experimental cycles is due to the higher dipolar interaction
in NWs array. A quick overview of the already reported theoretical
results further support these observations that the magnetic domains
and magnetization reversal inside an infinite NW and a NT of larger
diameters is not homogeneous {[}27,28,29,30{]}. In this work, this
is indirectly shown experimentally by combining the MFM and AGFM measurements.

\begin{figure}
\includegraphics[bb=30bp 80bp 360bp 580bp,clip,width=8.5cm]{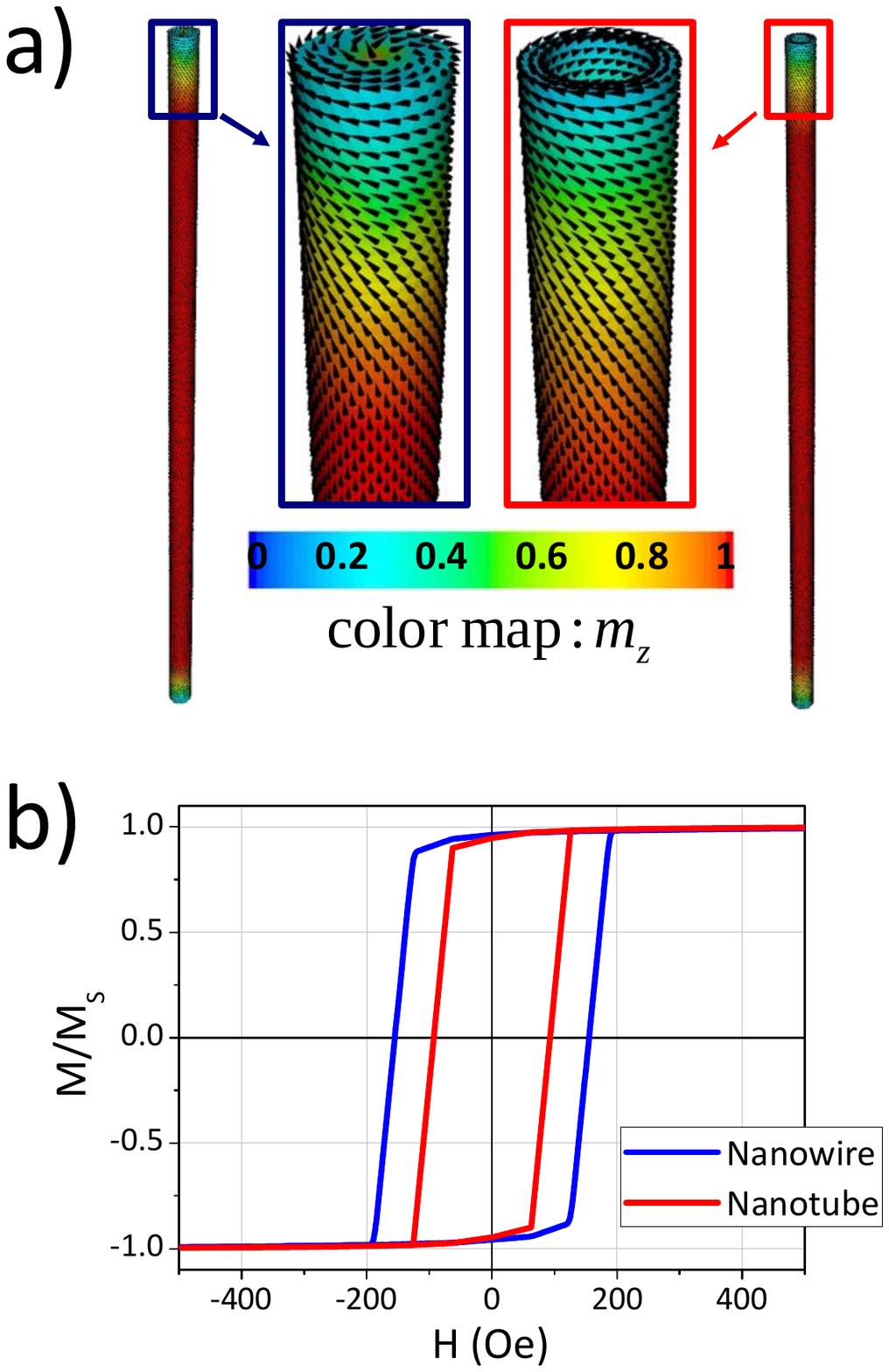} 

\caption{a) Magnetic moments distribution in an isolated NW and NT at zero
applied field after applying a saturating field along the revolution
axis. Colors encode mz component of the magnetization (along the revolution
axis). (b) Magnetization curves obtained for the isolated NW and the
NT, respectively. }
\end{figure}

The reasons why nanostructures of smaller dimensions are used to achieve
the high density storage media are two folds: first, to obtain a larger
density and,second, to avoid the inhomogeneous magnetic domains. The
nanostructures used here have negligible intrinsic switching field
distribution due to diameter distribution since they have larger diameters
{[}10{]}. The objective of using these dimensions here was to study
the difference of reversal mechanisms and magnetic interactions in
NWs and NTs. It is shown that even if the intrinsic SFD is not significantly
modified, it is still slightly broader compared to lower packing density
systems, especially for the NWs. This is the result of stronger dipolar
interactions and inhomogeneity of domains. For high density systems,
with smaller NT or NW diameter and low pitch, it is of prime importance
to have a deep insight into not only magnetic interactions but also
their intrinsic switching field which is strongly dependent upon the
diameter distribution. The next step would be the study of systems
with smaller diameters and reduced pitch where the present work will
serve as a reference.

\section{Conclusion}

The magnetization reversal process of Ni NWs and NTs arrays in PC
template have been investigated using MFM and AGFM. By comparing the
magnetization curves obtained from both techniques, it has been demonstrated
that they are complementary if one wants to get an insight of the
dipolar coupling and the magnetization reversal process. The presented
results helped us understanding their magnetization reversal. For
instance the mismatch of the magnetization curves in both the cases
reveals that the micromagnetic configurations inside the NWs and NTs
are not coherent. NWs array demonstrated stronger magnetic interactions
than the NTs arrays. These results may serve as a benchmark for comparing
the behavior of NWs and NTs and their use in various applications
accordingly. 
\begin{acknowledgments}
M.R. Tabasum is an Assistant Professor on leave from UET-RCET industrial
and manufacturing engineering (IME) department. Financial support
was provided by the Fédération Wallonie-Bruxelles (ARC 13/18-052 Supracryst)
and by the Belgian Federal Science Policy (IAP-PAI 7/05).\end{acknowledgments}

\end{document}